\documentclass[prd,floats,floatfix,nofootinbib,twocolumn]{revtex4}
\usepackage[dvips]{graphicx}
\usepackage{graphics}
\usepackage{dcolumn}
\usepackage{amssymb}
\usepackage{bm}
\usepackage{amsmath}
\usepackage{color}
\usepackage{epstopdf}

\def\spose#1{\hbox to 0pt{#1\hss}}

\def\lta{\mathrel{\spose{\lower 3pt\hbox{$\mathchar"218$}}
     \raise 2.0pt\hbox{$\mathchar"13C$}}}
\def\gta{\mathrel{\spose{\lower 3pt\hbox{$\mathchar"218$}}
     \raise 2.0pt\hbox{$\mathchar"13E$}}}
\newcommand{\be}{\begin{equation}}
\newcommand{\en}{\end{equation}}
\newcommand{\bea}{\begin{eqnarray}}
\newcommand{\ena}{\end{eqnarray}}

\begin{document}

\title{Strong Lensing
and Nonminimally Coupled Electromagnetism }

\author{Santiago E. P. Bergliaffa\footnote{sepbergliaffa@gmail.com},
Edson Elias de Souza Filho\footnote{ee\_souzafilho@outlook.com},
Rodrigo Maier\footnote{rodrigo.maier@uerj.br}
\vspace{0.5cm}}

\affiliation{Departamento de F\'isica Te\'orica, Instituto de F\'isica, Universidade do Estado do Rio de Janeiro,\\
Rua S\~ao Francisco Xavier 524, Maracan\~a,\\
CEP20550-900, Rio de Janeiro, Brazil\\
}


\date{\today}

\begin{abstract}
The lensing at large deflection angles caused by a
Schwarzschild black hole 
for the case of a nonminimal coupling between gravitation and electromagnetism is examined. We show that photons follow an effective geometry, which 
displays an effective photon sphere.  For the case in 
which the source,
lens and observer are aligned, so that relativistic Einstein rings are formed,
the dependence of the angular separation $\delta\theta$ between the first and second ring with the relevant coupling parameter is calculated. We argue that such a separation, which may be measured by telescopes that will be operative in the near future,
may set an upper and a lower limit for the coupling parameter. 

%
\end{abstract}
\maketitle
\section{Introduction}

The bending of light rays 
by a Schwarzschild black hole 
in the case in which the rays have an impact parameter close to $r=3M$ (corresponding to the photon sphere of the black hole) 
is an example of lensing at large deflection angles
\footnote{As opposed 
to lensing at small deflection angles (which may lead to multiple images) described in the weak gravitational field regime (sometimes called ``strong lensing''), see for instance \cite{matthias}.}. Such a situation was studied first in \cite{darwin}
\footnote{See also \cite{atkinson}.}, reexamined in\cite{perlick}-\cite{tsupko}, and analysed using a new
lens equation for the  
the deflection
angle in
\cite{ellis}.
It was shown in the latter reference that, considering the case where the observer, lens and source are aligned,
an infinite sequence of relativistic Einstein rings is obtained. In the misaligned case,
an infinite 
sequence of relativistic images (also called higher-order images)
is produced
on both sides of the optical axis
, as well as the primary and the
secondary images \cite{ellis}. The relativistic images are very much demagnified,
even in the case the source, the lens, and the
observer are perfectly or highly aligned\cite{tsupko2008}, although in the latter case the magnification is somewhat larger than the former. 
The analysis of lensing at large deflection angles was extended in \cite{vir2}, where 
the influence on the position of the relativistic images due to changes in the angular
source position as well as the lens-source and lens-observer distances was studied. 
The strong lensing efect has been since then studied in a variety of systems, such as
different types of black holes (see for instance
\cite{amarilla, shao1, shao2, papnoi}, and wormholes \cite{naoki}), among others.

It is important to stress that the abovementioned results were obtained under the assumption of a minimal coupling between electromagnetism and gravity.
However, 
%
%
more general couplings of the  electromagnetic field to gravitation are possible, as reviewed in \cite{goenner,hehl, balakin1}\footnote{For the generalization of the nonminimal coupling to include an axion see \cite{balakin2}.}. Among the multiple consequences of a nonminimal coupling (NMC), we can mention the following. The influence of such a coupling  on the dispersion relation for waves was studied in 
\cite{balakin3}. 
Exact pp-wave solutions for gravity and electromagnetism in  the NMC case were obtained in \cite{dereli1}. 
Black
holes solutions for such a system were 
presented in \cite{hoissen}, and reconsidered along with 
soliton solutions in \cite{balakin4},
and in \cite{dereli2},\cite{dereli3} with couplings of the type $f(R) F_{\mu\nu}F^{\mu\nu}$, while 
Einstein-Rosen bridges were obtained in \cite{balakin5}.
A static nonminimally coupled test magnetic field around a Schwarzschild
black hole 
was analyzed in 
\cite{petar}.
In a cosmological setting, 
nonminimal couplings between EM and gravity were applied to Bianchi I models with a magnetic field in \cite{balakin6}, while the influence of nonminimal couplings on the propagation of photons in the early universe was studied in \cite{balakin7}.

Since, as discussed in \cite{chu}, a possible NMC would not be probed by cosmological
propagation of light or solar system tests of General Relativity, we shall explore here the consequences of a NMC between 
gravitation and electromagnetism in the lensing at large deflection angles. In particular, we shall study 
the propagation of nonminimally coupled photons in the eikonal approximation in Schwarzschild's geometry. For those photons with
impact parameter 
close to the effective photon sphere, 
we shall obtain
the dependence of the 
the angular separation of the relativistic Einstein rings 
(formed when the observer, the lens, and the source are aligned)
with the parameter corresponding to the NMC. 
In Section II, the theoretical formulation that leads to an effective geometry for the photons, 
due to the NMC between gravitation and electromagnetism, is presented. In Section III, we examine the effective potential, and show how the position of the effective photon sphere depends of 
the coupling parameter. In Section IV the dependence 
with the 
coupling parameter of the main quantities related to lensing at large deflection angles:
the deflection angle $\alpha$,
the closest distance of approach $R_0$, and the angle $\theta$ of the image with respect to the optic axis is exhibited. Finally
we present the plot of the angular separation of the first two rings as a function of the coupling. We draw our final remarks in Section V.

%
%
%

\section{Nonminimal coupling and the effective geometry}
Let ${\cal S}_{p}$ be the action for a photon in a gravitational field. The equation of motion
for the electromagnetic field is then written as
\begin{eqnarray}
\label{eq1}
\frac{\delta {\cal S}_{p}}{\delta A_\mu}=0
\end{eqnarray}
where $A_\mu$ is the potential $4$-vector connected to the Faraday tensor $F_{\mu\nu}=A_{\nu,\mu}-A_{\mu,\nu}$ and the commas
denote usual partial differentiation. The minimally coupled part of ${\cal S}_{p}$ is given by the Maxwell
Lagragian:
\begin{eqnarray}
\label{eq2}
{\cal S}_{0}=-\frac{1}{4}\int \sqrt{-g} F_{\mu\nu}F^{\mu\nu} d^4x.
\end{eqnarray} 
The non-minimally coupled sector
%
is described by the action
\begin{eqnarray}
\label{eq3}
\nonumber
{\cal S}_{1}= \int \sqrt{-g} (\gamma_1 R F_{\mu\nu}F^{\mu\nu}+\gamma_2 R_{\mu\nu}F^{\mu}_{~~\beta}F^{\nu\beta}\\
+\gamma_3 R_{\mu\nu\beta\sigma}F^{\mu\nu}F^{\beta\sigma}
)d^4x,
\end{eqnarray}
where $\gamma_i$ ($i=1, 2, 3$) are coupling coefficients,
$R_{\mu\nu\beta\sigma}$ is the Riemann tensor, $R_{\mu\nu} \equiv R^\alpha_{~~\mu\alpha\nu}$, $R\equiv g^{\mu\nu}R_{\mu\nu}$
and $\nabla_\mu$ is the covariant derivative built with the Christoffel symbols. Therefore, assuming that ${\cal S}_{p}={\cal S}_{0}+{\cal S}_{1}$, the equations of motion for the electromagnetic field can be rewritten as
\begin{eqnarray}
\label{eq4}
\nabla_\mu F^{\mu\nu}+\frac{\delta {\cal S}_{1}}{\delta A_\nu}=0.
\end{eqnarray}
%
The corresponding equations of motion are
\footnote{These equations (with fixed values for the $\gamma$s) can be obtained by considering QED vacuum polarization effects \cite{drummond}.}:
\begin{eqnarray}
\label{eq5}
\nonumber
\nabla_{\mu} F^{\mu\nu}+2\nabla_\mu [2\gamma_1 R F^{\mu\nu}+\gamma_2(R^\mu_{~~\beta}F^{\beta\nu}-R^\nu_{~~\beta}F^{\beta\mu})\\
+2\gamma_3 R^{\mu\nu}_{~~~\beta\sigma}F^{\beta\sigma}]=0.
\end{eqnarray}
Restricting to vacuum spacetimes which satisfy Einstein's field equations $R_{\mu\nu}=0$, 
equation (\ref{eq5}) reads
\begin{eqnarray}
\label{eq51}
\nabla_{\mu} F^{\mu\nu}+4\gamma_3\nabla_\mu (
 R^{\mu\nu}_{~~~\beta\sigma}F^{\beta\sigma})=0.
\end{eqnarray}
The electromagnetic tensor obeys also the Bianchi identity:
\begin{eqnarray}
\label{eq54}
\nabla_{\alpha}F_{\mu\nu}+\nabla_{\mu}F_{\nu\alpha}+\nabla_{\nu}F_{\alpha\mu}=0,
\end{eqnarray}
To study the lensing of rays governed by
Eqs. \eqref{eq51} and \eqref{eq54} in a Schwarzschild spacetime, we shall 
use the eikonal approximation, 
%
in which the test electromagnetic field is given by
\begin{eqnarray}
\label{eq52}
F_{\mu\nu}=f_{\mu\nu}e^{i\theta},
\end{eqnarray}
where 
the phase $\theta$ is a very rapidly-varying function 
(on scales much lower than the curvature scale, and much higher than 
the Compton wavelength of the electron)
compared to the amplitude 
$f_{\mu\nu}$. 
By defining $k_\mu:=\theta_{,\mu}$, equation (\ref{eq51}) yields
\begin{eqnarray}
\label{eq53}
k_\mu f^{\mu\nu}+4\gamma_3 k_\mu R^{\mu\nu}_{~~~\beta\sigma}f^{\beta\sigma}=0.
\end{eqnarray}  
On the other hand, we obtain from the Bianchi identities \eqref{eq54},
\begin{eqnarray}
\label{eq55}
k_{\alpha}f_{\mu\nu}+k_{\mu}f_{\nu\alpha}+k_{\nu}f_{\alpha\mu}=0.
\end{eqnarray}
Contracting Eq. (\ref{eq55}) with $k^\alpha$ and using Eq. (\ref{eq54}) we obtain
\begin{eqnarray}
\label{eq56}
k^2f_{\mu\nu}+4\gamma_3 k_\alpha(k_\mu R^\alpha_{~~\nu\beta\sigma}-k_\nu R^\alpha_{~~\mu\beta\sigma})f^{\beta\sigma}=0.
\end{eqnarray}
From the 
Schwarzschild metric given in  standard coordinates $x^\alpha=(t,r,\theta,\phi)$,
\begin{eqnarray}
\label{eq57}
\nonumber
ds^2=\Big(1-\frac{2M}{r}\Big)dt^2-\Big(1-\frac{2M}{r}\Big)^{-1}dr^2\\
-r^2(d\theta^2+\sin^2{\theta}d\phi^2),
\end{eqnarray}
%
an orthonormal tetrad as the base of $1$-forms 
\begin{eqnarray}
\label{eq58}
\Theta^A:=e^A_{~\alpha} dx^\alpha
\end{eqnarray}
can be defined,
where
\begin{eqnarray}
\label{eq59}
e^A_{~~\alpha}\rightarrow {\rm diag}\Big(\sqrt{1-\frac{2M}{r}}, ~1\Big/\sqrt{1-\frac{2M}{r}}, ~r, ~r\sin{\theta}\Big).
\end{eqnarray}
Hence the line element can be written as
\begin{eqnarray}
\label{eq60}
ds^2=\eta_{AB}\Theta^A\Theta^B,
\end{eqnarray}
where $\eta_{AB}$ is the usual Minkowski metric of the tangent space. Let $Q^{AB}$ and $\Omega^{AB}$ be two bivectors defined as
\begin{eqnarray}
\label{eq61}
Q^{AB}&=&\sqrt{-g^{tt}g^{rr}}(e^{A}_{~~t}e^{B}_{~~r}-e^{A}_{~~r}e^{B}_{~~t}),\\
\Omega^{AB}&=&\sqrt{g^{\theta\theta}g^{\phi\phi}}(e^{A}_{~~\theta}e^{B}_{~~\phi}-e^{A}_{~~\phi}e^{B}_{~~\theta}),
\end{eqnarray}
so that the Riemann tensor can be written as
\begin{eqnarray}
\label{eq62}
\nonumber
R^\alpha_{~~\nu\beta\sigma}=\frac{M}{r^3}[\delta^\alpha_{~\beta}g_{\nu\sigma}-\delta^\alpha_{~\sigma}g_{\nu\beta}~~~~~~~~~~~~~~~~~~~~\\
~~~~~~~~~~~~~~~~~~+3(Q^\alpha_{~~\nu}Q_{\beta\sigma}-\Omega^\alpha_{~~\nu}\Omega_{\beta\sigma})],
\end{eqnarray}
where $Q^{\alpha\beta}\equiv Q^{AB} e_{A}^{~~\alpha} e_{B}^{~~\beta}$ and  $\Omega^{\alpha\beta}\equiv W^{AB} e_{A}^{~~\alpha} e_{B}^{~~\beta}$, with $e^{A}_{~~\alpha}e_{A}^{~~\beta}\equiv \delta^{\alpha}_{~~\beta}$. Substituting
Eq.(\ref{eq62}) in Eq.(\ref{eq56})
we obtain
\begin{eqnarray}
\label{eq63}
\nonumber
\Big(1+\frac{8M\gamma_3}{r^3}\Big)k^2f_{\mu\nu}~~~~~~~~~~~~~~~~~~~~~~~~~~~~~~~~~~~~~~\\
\nonumber
+\frac{12M\gamma_3}{r^3}k_\alpha\Big[k_\mu(Q^\alpha_{~~\nu}
Q_{\beta\sigma}-\Omega^\alpha_{~~\nu}\Omega_{\beta\sigma})~~~~~~~~~~~~~\\
-k_\nu(Q^\alpha_{~~\mu}
Q_{\beta\sigma}-\Omega^\alpha_{~~\mu}\Omega_{\beta\sigma})\Big]f^{\beta\sigma}=0.
\end{eqnarray}
Defining the scalars
\begin{eqnarray}
\label{eq64}
q:=Q_{\alpha\beta}f^{\alpha\beta}~~{\rm and}~~\omega:= \Omega_{\alpha\beta}f^{\alpha\beta},
\end{eqnarray}
together with the vectors
\begin{eqnarray}
\label{eq65}
l^\mu:=k_\alpha Q^{\alpha\mu}~~{\rm and}~~m^\mu:= k_\alpha \Omega^{\alpha\mu},
\end{eqnarray}
equation (\ref{eq63}) turns into
\begin{eqnarray}
\label{eq66}
k^2f_{\mu\nu}+\frac{\Delta}{2}[q(k_\mu l_\nu -k_\nu l_\mu )-\omega(k_\mu m_\nu -k_\nu m_\mu )]=0,
\end{eqnarray}
where
\begin{eqnarray}
\label{eq67}
\Delta=\frac{24M\gamma_3}{r^3+8M\gamma_3}.
\end{eqnarray}
By contracting Eq. (\ref{eq66}) with $Q^{\mu\nu}$ we obtain
%
\begin{eqnarray}
\label{eq68}
(k^2+l^2 \Delta)q=0,
\end{eqnarray}
%
%
%
where $l^2\equiv l_\mu l^\mu$. Since $q\neq 0$, the modified light cone
follows from $k^2+l^2\Delta=0$ \cite{drummond}, or
\begin{eqnarray}
\label{eq69}
(1-\Delta)(k_tk^t+k_rk^r)+k_\theta k^\theta +k_\phi k^\phi =0.
\end{eqnarray}
The latter leads to the effective geometry
$\tilde{g}_{\mu\nu}$
(see \cite{barcelo} for a review), in such a way that the line element for the light rays in the non-minimally coupled case is given by 
\footnote{A similar expression was obtained in \cite{chu}.}
\begin{eqnarray}
\label{eq70}
\nonumber
d\tilde{s}^2=\tilde{g}_{\mu\nu}dx^\mu dx^\nu=\Big(1-\frac{1}{R}\Big)(1-\Delta)dT^2~~~~~~~~~~\\
-\Big(1-\frac{1}{R}\Big)^{-1}(1-\Delta)dR^2
-R^2(d\theta^2 +\sin{\theta}d\phi^2),
\end{eqnarray}
with
\begin{eqnarray}
\label{eq71}
T=\frac{t}{2M},~~R=\frac{r}{2M},~~
\Delta=\frac{3\Gamma_3}{R^3+\Gamma_3},
\end{eqnarray}
and ${\Gamma}_3\equiv \gamma_3/M^2$.
Notice that the zero 
of 
$1-\Delta$, given by 
 $R_e=(2\Gamma_3)^{1/2}$ is such that $R_e<<1$, since we expect that 
 $\Gamma_3<<1$.

\section{The Effective Potential}

Let us consider a photon which follows a null geodesic in the effective metric defined by Eq.(\ref{eq70}).
The first integral of the equations of motion reads
\begin{eqnarray}
\label{t3}
\nonumber
f({R})\Big(\frac{dT}{d\lambda}\Big)^2-\frac{1}{g(R)}\Big(\frac{dR}{d\lambda}\Big)^2-R^2 \Big(\frac{d\theta}{d\lambda}\Big)^2~~~~~~\\
-
R^2\sin^2{\theta} \Big(\frac{d\phi}{d\lambda}\Big)^2=0,
\end{eqnarray}
where $\lambda$ is an affine parameter and  
%
%
%
%
\begin{eqnarray}
\label{t2}
f(R)&=& \Big( 1-\frac{1}{R}\Big)
(1-\Delta),
\\
g(R)&=& \Big( 1-\frac{1}{R}\Big)
(1-\Delta)^{-1}.
\end{eqnarray}
The symmetries of the background metric guarantee the existence of 
two constants of motion, namely $E$ and ${\cal J}$,
given by
\begin{eqnarray}
\label{t4}
E=f(R)\frac{dT}{d\lambda}~~{\rm and}~~{\cal J}=R^2\frac{d\phi}{d\lambda}.
\end{eqnarray}
Therefore, the above first integral can be rewritten as
\begin{eqnarray}
\label{t7}
\frac{f(R)}{g(R)}\Big(\frac{dR}{d\lambda}\Big)^2+V_{\rm eff}(R)=E^2.
\end{eqnarray}
where
\begin{eqnarray}
\label{t8}
V_{\rm eff}(R)\equiv f(R)\frac{{\cal J}^2}{R^2}
\end{eqnarray}
is the effective potential, the maximum of which defines the photon sphere
$R_p$.
Feeding Eq. (\ref{t8}) with Eq.  (\ref{t2}), it is straightforward to see that the solutions of 
\begin{figure}[tbp]
\includegraphics[width=8.5cm,height=6cm]{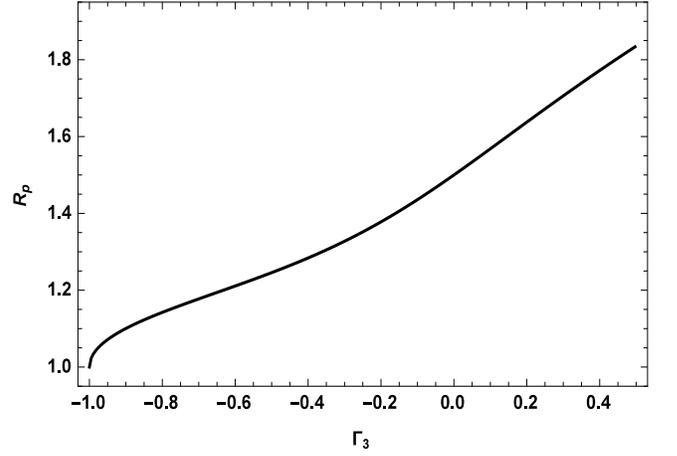}
\caption{$R_p$ as a function of $\Gamma_3$.}
\end{figure}
%
%
%
\begin{eqnarray}
\label{t9}
\frac{dV_{eff}}{dR}\Big|_{R_p}=0
\end{eqnarray}
are given by
\begin{eqnarray}
\label{t10}
\Gamma_{3\pm} =\Big[ \frac{(12-11R_p)\pm 3\sqrt{24+R_p(17R_p-40)}}{4(2R_p-3)}\Big]R_p^3.
\end{eqnarray}
%
It follows that 
%
\begin{eqnarray}
\label{lim}
\lim_{R_p\rightarrow 1}\Gamma_{3+}=-1~~{\rm and}~~\lim_{R_p\rightarrow 1}\Gamma_{3-}=\frac{1}{2}.
\end{eqnarray}
We shall 
consider here only the ``$+$'' branch, since the
``$-$'' branch does not allow small values of $\Gamma_3$.

For $-1<\Gamma_3< 1/2$ the effective potential $V_{eff}$ looks like one in the $\Gamma_3=0$ case. In particular, it exhibits only one global maximum $R_{p+}$, which is the effective photon sphere.
%
%
In order to simplify our notation, instead of $R_{p+}$ we will denote
the global maximum just $R_p$ in the following.
The variation of $R_p$ with $\Gamma_3$ is shown in Fig. 1.

\section{Strong Field Lensing:
minimal vs. nonminimal coupling.}
\label{strong}
We have seen in the previous section that the effective geometry given by Eq. (\ref{eq70}) generates a modified light cone on which
photons in the eikonal approximation propagate. Standard calculations (see for instance
\cite{wein})
using the metric (\ref{eq70}) instead of the Schwarzschild metric 
lead to 
the following expression for the deflection angle:
\begin{eqnarray}
\label{eq73}
\alpha=2\int^{\infty}_{r_0}\frac{\sqrt{1-\Delta} dr}{r\sqrt{\frac{r^2}{r_0^2}\Big(1-\frac{2M}{r_0}\Big)\frac{(1-\Delta_0)}{(1-\Delta)}-\Big(1-\frac{2M}{r}\Big)}}-\pi,
\end{eqnarray}
with impact parameter
\begin{eqnarray}
\label{ip}
{\cal J}=r_0 \Big[\Big(1-\frac{2M}{r_0}\Big)(1-\Delta_0)\Big]^{-\frac{1}{2}}.
\end{eqnarray}
Here $r_0$ is the closest distance of approach. For the purpose of computation, it is useful to use the rescaled coordinates  introduced in Eq.(\ref{eq71}).
In this case, the deflection angle $\alpha$ can be written as
\begin{eqnarray}
\label{eq75}
\alpha=2\int^{\infty}_{R_0}\frac{\sqrt{1-\Delta} dR}{R\sqrt{\frac{R^2}{R_0^2}\Big(1-\frac{1}{R_0}\Big)\frac{(1-\Delta_0)}{(1-\Delta)}-\Big(1-\frac{1}{R}\Big)}}-\pi,
\end{eqnarray}
and
\begin{eqnarray}
\label{ipn}
{\cal J}=2M R_0 \Big[\Big(1-\frac{1}{R_0}\Big)(1-\Delta_0)\Big]^{-\frac{1}{2}}.
\end{eqnarray}
All the formulas given here reduce to the minimally coupled case when $\Delta = 0$.
\footnote{In the weak field limit the integrand of (\ref{eq75}) can be written as a power series of $M/r$ and $M/r_0$.
Following the standard calculation presented in \cite{wein}, it can be shown that the correction due to $\Gamma_3$
is of the order of ($M^3/r^3$, $M^3/r_0^3$). Hence, it is negligible in the weak field regime.}

In this paper we are going to consider a simple lens equation for
asymptotically flat spacetimes
originally introduced by K. S. Virbhadra and G. F. R. Ellis \cite{ellis}.
\footnote{The more general equation presented in \cite{bozza2008,aazami} reduces to the one used here
for the case in which the observer, lens and source are aligned.}
The lens diagram is depicted in Fig. 3. 
\begin{figure}[tbp]
\centering
\begin{center}
\includegraphics[width=6cm,height=6.5cm]{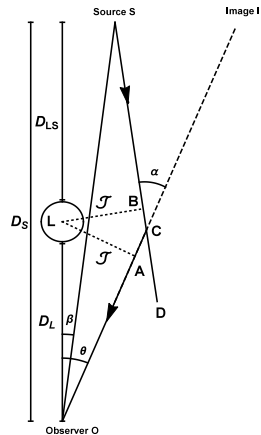}
\caption{Lens diagram. $\alpha$ is the Einstein deflection angle,
$\beta$ and $\theta$
are the angular position of the source and of its image with respect to the optic axis -- defined by the line joining 
the observer $O$ and
the lens $L$.
$\overline{LA}$ and $\overline{LB}$ -- which furnish the impact parameter ${\cal J}$ -- are perpendicular lines to $\overline{SD}$ and $\overline{OI}$,
respectively. Finally, while $D_{LS}$ is the lens-source distance, $D_S$
and $D_L$ are the distances of the source and the lens
to the observer.}
\end{center}
\end{figure}
Let $\alpha$ be the Einstein deflection angle.
If $\beta$ and $\theta$
are the angles of the source and its image with respect to the optic axis -- defined by the line joining 
the observer $O$ and
the lens $L$ --
the lens equation reads 
\begin{eqnarray}
\label{eqSL1}
\tan{\beta}=\tan{\theta}-\sigma,
\end{eqnarray}
where
it is assumed that both $O$ and $S$ are located
far from the lens, in asymptotically flat 
regions,
and
\begin{eqnarray}
\label{eqSL2}
\sigma=\frac{D_{LS}}{D_S}[\tan{\theta}+\tan{(\alpha-\theta)}].
\end{eqnarray}
Furthermore,
from the lens diagram we obtain that
\begin{eqnarray}
\label{eqSL3}
{\cal J}=D_L \sin{\theta}.
\end{eqnarray} 

In order to better understand the gravitational lensing due to the nonminimal coupling,
it is useful to bear in mind the results of the minimally coupled case \cite{ellis}.
It is well-known that 
for rays travelling through regions in which a spherically symmetric field gravitational field is weak an Einstein ring is formed when the source,
lens and observer are aligned. Otherwise, a pair of images -- usually called
primary and secondary -- of opposite parities is formed. Photons travelling close to the photon sphere ({\it i.e.} in the strong field regime), 
may go around the lens once, twice or many times. When the source,
lens and observer are aligned, an infinite number of relativistic Einstein rings 
are formed due to the 
bending of light rays larger than $2\pi$. For the case of misaligned components, an infinite sequence of relativistic images on both sides of the optic axis is obtained.
We shall restrict here to the case in which 
the source,
lens and observer are aligned in order to infer how the NMC affects the 
strong lensing. To compare our results
with those in the literature, we assume that the lens 
is the supermassive black hole at the center of our Galaxy \cite{ellis}, so that the parameters are:
$D_L=8.5{\rm kpc}$, $M=2.8\times 10^6M_{\odot}$, $D_S=2D_{LS}$,
where $M_{\odot}$ is the solar mass. 
%

To evaluate the position of the relativistic Einstein
rings we follow two different numerical procedures. The first one consists of the following steps:
  \begin{itemize}
        \item[1.i] 
        A finite number of pairs $(\Gamma_{3},R_p)$
        is evaluated from Eq. (\ref{t10}), in the  domain 
$-1\leq \Gamma_{3} \leq 0.5$
;
\item[1.ii] For each value of $\Gamma_{3}$ a 
finite
number of values of  
$R_0$, defined by 
$R_0(j)=R_p$
$+j\epsilon$ is generated, where $\epsilon$
is a sufficiently small  
increment and $j$ an 
integer;
\item[1.iii] Given $R_0(j)$ in (1.ii), we evaluate
from Eq.(\ref{eq75})  the corresponding values of $\alpha$ (for a fixed $\Gamma_{3}$).
    \end{itemize}
The outcome of this first part is analogous to that of \cite{bozza}, and furnishes the plot of $\alpha$ as a function of $R_0$
for a fixed value of $\Gamma_{3}$ (cf. black curve in Fig. 4, built with $\Gamma_3=0$).
\begin{figure}
\includegraphics[width=8.5cm,height=6cm]{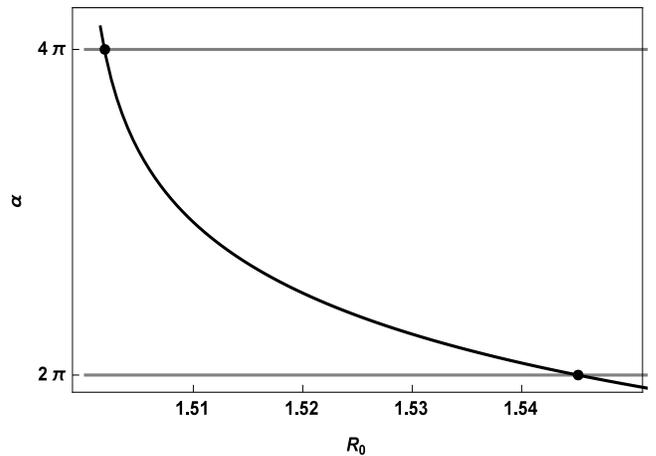}
\caption{The deflection angle $\alpha$ as a function of $R_0$ for $\Gamma_{3}=0$. The black curve is obtained in the first part of our numerical procedure (see text)
where we use the integral in Eq. (\ref{eq75}). The gray curves are the product of the second part of our numerical procedure ,
where we use the lens equation (\ref{eqlr}).
The points of intersections between the black and gray curves furnish the angular deflections
of the first and second relativistic Einstein rings.  }
\end{figure}

The second part of the procedure refers to the lens equation and is rather more involved:
\begin{itemize}
        \item[2.i] Eqs. (\ref{ipn}) and (\ref{eqSL3}) are evaluated at the  distance of
        closest
approach, namely the photon sphere, thus obtaining
\begin{eqnarray}
\label{x01}
D_L \sin{\theta_p}\equiv 2M R_p \Big[\Big(1-\frac{1}{R_p}\Big)(1-\Delta_p)\Big]^{-\frac{1}{2}}.
\end{eqnarray}
Feeding (\ref{x01})
with each pair $(\Gamma_{3},R_p)$ from (1.i), 
the corresponding $\theta_p$
is evaluated;

\item[2.ii] For each numerical value of $\Gamma_{3}$ a 
finite
number of values of  
$\theta_0$, defined by $\theta_0(j)=\theta_p$
$+j\tilde{\epsilon}$ is generated, where $\tilde\epsilon$
is a 
small
increment and $j$ 
an integer;

\item[2.iii] Inserting 
the $\theta_0(j)$ from 
(2.ii) in the lens equation
\begin{eqnarray}
\label{eqlr}
\tan{\theta}=\frac{D_{LS}}{D_S}[\tan{\theta}+\tan{(\alpha-\theta)}]
\end{eqnarray} 
the 
corresponding $\alpha(j)$ are obtained;

\item[2.iv] Using Eqs.(\ref{ipn}) and (\ref{eqSL3}) so that
\begin{eqnarray}
\label{x0}
D_L \sin{\theta_0}\equiv 2M R_0 \Big[\Big(1-\frac{1}{R_0}\Big)(1-\Delta_0)\Big]^{-\frac{1}{2}},
\end{eqnarray}
all the remaining $R_0(j)$ corresponding to $\theta_0(j)$
are evaluated.
    \end{itemize}
    As in the first part, this procedure allows to 
plot $\alpha$ as a function of $R_0$
for a fixed value of $\Gamma_{3}$ (see gray curves in Fig. 4).
The points of intersection between the curve generated in the first part and the curves 
generated in the second part furnish the angular deflections
of the first and second relativistic Einstein rings,
given respectively by 
$2\pi+33.80\;\mu$as and 
$4\pi+33.75\;\mu$as \cite{ellis}.

We shall show next how the nonminimal coupling 
changes the values of $\alpha$
obtained for $\Gamma_3=0$.
In Fig. 5 the  effective deflection angle of the first (upper panel) and second (lower panel)
relativistic Einstein rings are shown when the nonminimimal coupling is present. 
These plots show that, depending on the sign of $\Gamma_3$, the effective deflection angle can be smaller or larger than that of the minimally coupled case. 
For completeness, 
the  distance of closest approach $R_0$ as a function of $\Gamma_3$ is shown in Fig 6
for both rings. 
\begin{figure}
\includegraphics[width=8.5cm,height=6cm]{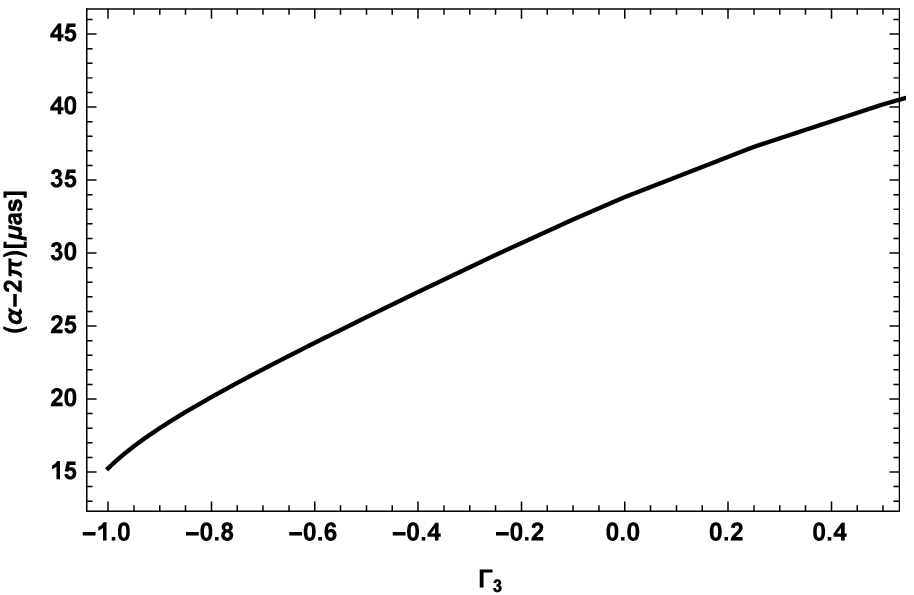}\\
\includegraphics[width=8.5cm,height=6cm]{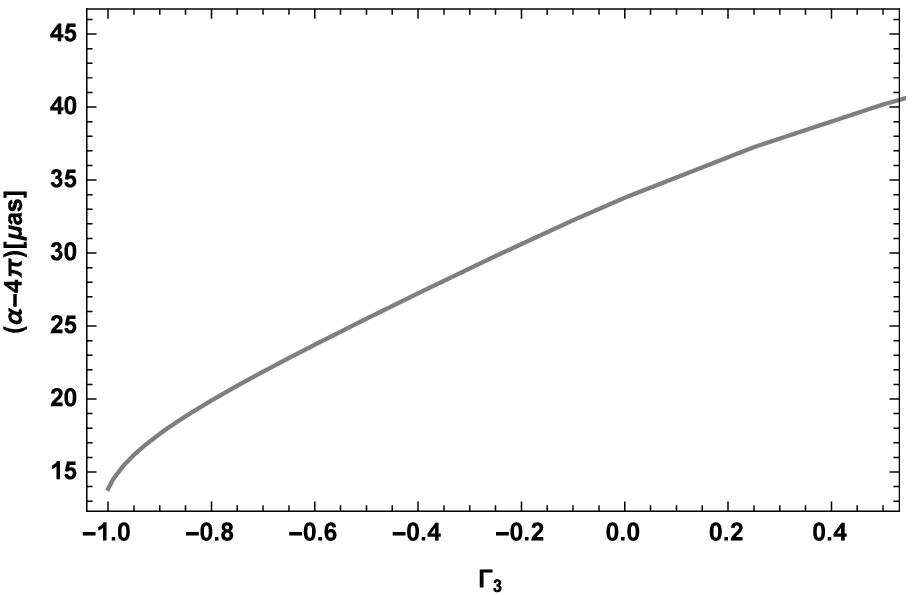}
\caption{Effective deflection angles as a function of the nonminimal coupling parameter
of the first (upper panel) and second (lower panel)
relativistic Einstein rings. The results for the minimally coupled case are recovered when $\Gamma_3=0$ \cite{ellis}.}
\end{figure}
\begin{figure}
\includegraphics[width=8.5cm,height=6cm]{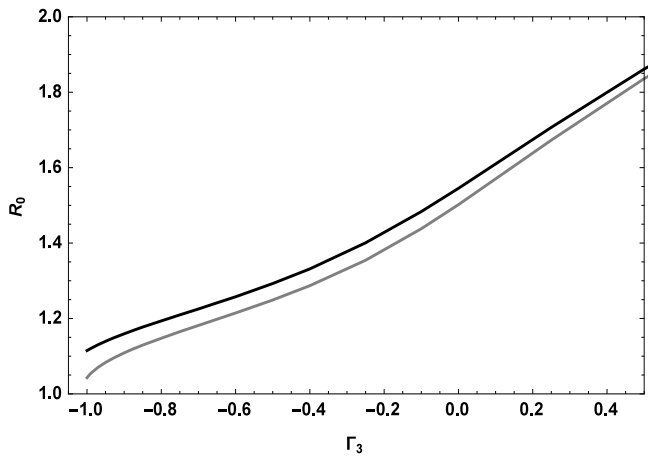}
\caption{The closest distance of approach $R_0$ as a function of $\Gamma_3$ for the first (black curve) and second (gray curve)
relativistic Einstein rings.}
\end{figure}

In Fig. 7 the angular position of the first (upper panel) and second (lower panel)
relativistic Einstein rings
as a function of $\Gamma_3$
is displayed. At this stage it is useful to distinguish such angles 
using a different notation:
$\theta_{2\pi(4\pi)}$ denotes the 
angular position of the first (second) relativistic
Einstein ring,
%
%
%
and define the separation
\begin{eqnarray}
\delta \theta = \theta_{2\pi}- \theta_{4\pi}.
\end{eqnarray}
\begin{figure}
\includegraphics[width=8.5cm,height=6cm]{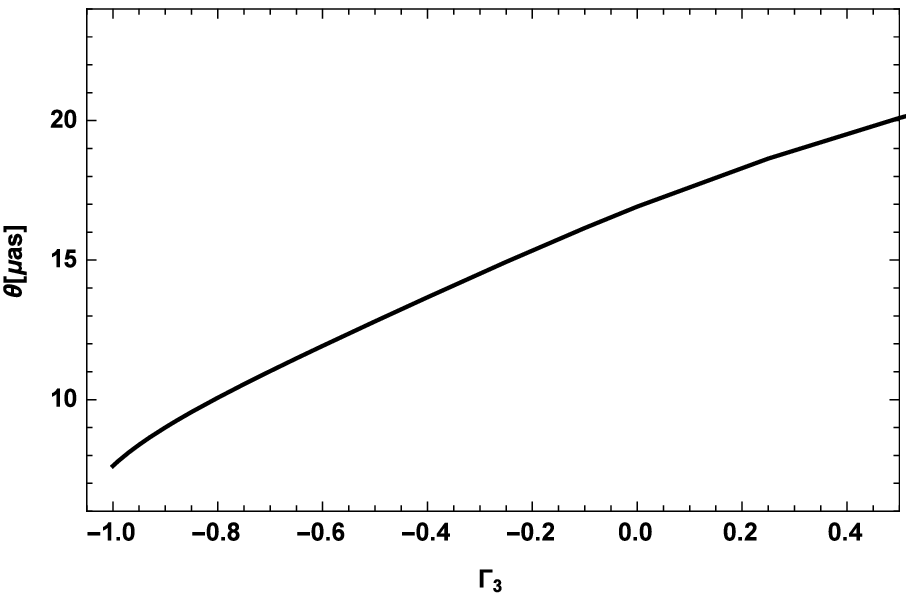}\\
\includegraphics[width=8.5cm,height=6cm]{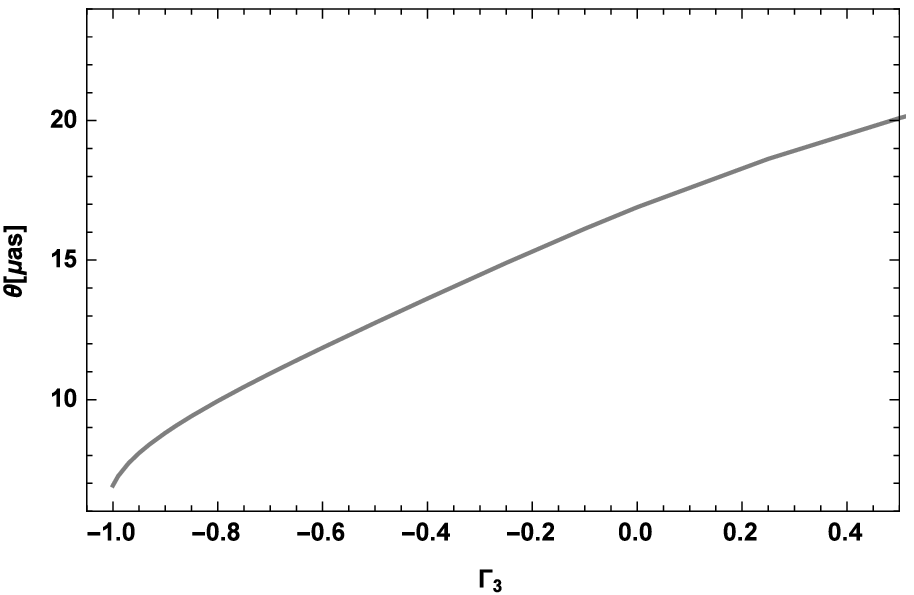}
\caption{The angular positions of the first (upper panel) and second (lower panel)
relativistic Einstein rings as a function of $\Gamma_3$.}
\end{figure}
Fig. \ref{deltatheta} displays $\delta \theta$ as a function of $\Gamma_3$. 
%
It is worth noting that the result for the minimally coupled case 
(namely, 
$\delta\theta\approx 2\times 10^{-2}\mu$as \cite{ellis}) is
recovered 
in the limit $\Gamma_3\rightarrow 0$.  
\begin{figure}
\label{deltatheta}
\includegraphics[width=8.5cm,height=6cm]{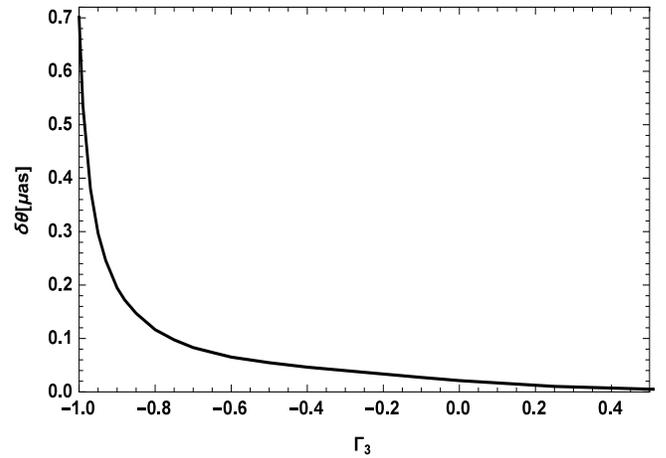}
\caption{The separation between the first and second
relativistic Einstein rings as a function of $\Gamma_3$.
This separation falls rapidly as $\Gamma_3$ increases, reaching
the order of $5\times 10^{-3}\mu{\rm as}$ for $\Gamma_3 = 0.5$.}
\end{figure}

\section{Final Remarks}

We have examined strong deflection effects taking into account a 
nonminimal coupling between gravitation and electromagnetism.
Assuming a Schwarzschild background, it was shown that the motion of photons 
in the eikonal limit 
is governed by 
an effective metric. 
The associated effective potential displays an effective photon sphere, the radius of which depends of $\Gamma_3$.
We have analized here a particular configuration that displays strong field lensing effects in which
the observer, the lens (taken as the supermassive black hole at the center of our Galaxy), and the source are aligned
so that an infinite number of relativistic Einstein rings are formed. 
Using the lens equation introduced in \cite{ellis}, we evaluated in the strong field lensing regime the dependence with $\Gamma_3$ 
of the deflection angle $\alpha$,
the closest distance of approach $R_0$, and the angular position $\theta$ of the image with respect to the optical axis.
Our results show that the angular separation $\delta\theta$ between the first and second Einstein rings
can be as high as approx. 0.7$\mu$as (for $\Gamma_3=-1$), and falls rapidly as the coupling parameter $\Gamma_3$ increases, reaching
the order of $10^{-3}\mu{\rm as}$ for $\Gamma_3 = 0.5$.

It is worth noting that the highest resolution telescope
available today
(the {\it Event Horizon Telescope}) \cite{kaz}
has a resolution of the order of $25 ~\mu{\rm as}$, \emph{i.e.} roughly $30$ times larger 
than the separation between the first and second Einstein rings
for $\Gamma_3=-1$. Hence, it
is not capable of detecting  
the maximum angular separation between two consecutive relativistic Einstein rings 
predicted by our results. If we assume that future observations yield  a result that will not be very different from that of the minimally coupled case (which is of the order of $2\times 10^{-2}\mu$as), at least a resolution of the order of
$10^{-2}\mu$as would be needed to set limits on $\Gamma_3$ using the results presented here. Such a resolution may be achieved by future instruments, in particular by the Millimetron Space Observatory \cite{kaz} (which 
may have a resolution of approx.
50 narcsec at $\lambda=$ 0.345 mm.). We should also point out that although the magnification for higher-order images
is very small in the vacuum case, it becomes significantly larger in the presence of plasma\cite{tsupko,tsupko2}.
Another important point is that a measure of $\delta\theta$ with its corresponding error would permit to set both an upper and a lower limit for $\Gamma_3$. This would an improvement with respect to the current situation, in which only upper limits are available \cite{prasanna}.

The results presented here, based on the aligned configuration, show that it may be feasible in the future to use the separation of the relativistic rings to set limits on $\Gamma_3$. 
The theoretical estimates presented here may improve if other configurations for the source-lens-observer system are studied. 
In particular,  values 
different from $1/2$
for the ratio
$D_{LS}/D_S$ could be considered, 
as well as misaligned configurations.  
It would also be of interest
to study how variations in the angular position of the source (together with changes in the lens-source distance)
would furnish modifications -- due to the nonminimal coupling -- in the angular separations between any two 
relativistic images.
Finally, the analysis should 
be extended to the more realistic case of a Kerr black hole
In fact, as discussed in 
\cite{wong},
precise measurements
of the photon ring and even its subrings in this case are feasible using interferometry.
We shall examine these points in future work.

\end{document}